\documentclass[doublespacing]{elsart}

\usepackage{icarus}

\usepackage{natbib}

\bibpunct{(}{)}{;}{a}{,}{,}

\usepackage{graphicx}

\begin{document}

\begin{frontmatter}



\title{Albedos and diameters of three Mars Trojan asteroids}


\author[steward]{David E. Trilling}, 
\author[apl]{Andrew S. Rivkin},
\author[steward]{John A. Stansberry},
\author[cfa]{Timothy B. Spahr},
\author[steward,crudo]{Richard A. Crudo}, and
\author[davies]{John K. Davies}

\address[steward]{Steward Observatory, The University
of Arizona, 933 N. Cherry Avenue, Tucson, AZ 85721}
\address[apl]{Applied Physics Laboratory, Johns Hopkins University, 11100 Johns Hopkins Road, Laurel, MD 20723}
\address[cfa]{Harvard-Smithsonian Center for Astrophysics, 60 Garden Street, Cambridge, MA 02139}
\address[crudo]{Present address: Department of Physics, University of Connecticut, U-3046, 2152 Hillside Road, Storrs, CT 06269} 
\address[davies]{UK Astronomy Technology Centre, Blackford Hill, Edinburgh, EH9 3HJ, UK}



%
%
%
%
%


\end{frontmatter}



\begin{flushleft}
\vspace{1cm}
Number of pages: \pageref{lastpage} \\
Number of tables: \ref{lasttable}\\
Number of figures: \ref{lastfig}\\
\end{flushleft}


\begin{pagetwo}{Albedos and diameters of three Mars Trojans}

David E. Trilling \\
Steward Observatory \\
The University of Arizona \\
933 N. Cherry Avenue \\
Tucson, AZ 85721 \\
\\
Email: trilling@as.arizona.edu \\
Phone: (520) 626-1600 \\
Fax: (520) 621-9555

\end{pagetwo}

\begin{abstract}

We observed the Mars Trojan asteroids (5261) Eureka and
(101429) 1998~VF$_{31}$ and the candidate Mars Trojan
2001~FR$_{127}$ at 11.2~and
18.1~microns using Michelle on the Gemini North telescope.
We derive diameters of 
1.28, 0.78, and $<$0.52~km, respectively, with
corresponding geometric visible
albedos of 0.39, 0.32, and $>$0.14.
The albedos for Eureka and 1998~VF$_{31}$
are consistent with the taxonomic
classes and compositions (S(I)/angritic and
S(VII)/achrondritic, respectively) and implied
histories presented in a companion paper
by Rivkin et al.
Eureka's surface likely has a relatively high thermal inertia,
implying a thin regolith that is consistent
with predictions and the small size that
we derive.

\end{abstract}

\begin{keyword}
TROJAN ASTEROIDS\sep INFRARED OBSERVATIONS\sep REGOLITHS
\end{keyword}


\section{Introduction}

It has been known for nearly a century that a large population of
asteroids, known as Trojan asteroids, exists in a 1:1 resonance
with Jupiter.
Neptune is known to have five Trojan asteroids,
but the only rocky planet with
known Trojan asteroids is Mars.
At present, the
number of confirmed Mars Trojan asteroids
is four, with a handful of other candidates.
All but one orbit in the L5 (trailing) zone, with 1 in the L4 (leading)
zone.

Mars Trojans can be dynamically stable over the
age of the Solar System at inclinations
between 12~and 40~degrees
\citep{te99,scholl05};
all known and candidate Mars
Trojans reside in this dynamically stable
region of phase space and may therefore be
primordial objects.
If so, these bodies represent leftover
planetesimals from the formation of Mars.
It is therefore interesting that
there is evidence that suggests
a diverse history for these bodies.
\citet{rivkin03}
obtained visible spectra of the three largest Mars Trojans ---
(5261)~Eureka, (101429) 1998~VF$_{31}$, and (121514) 1999~UJ$_{7}$ ---
and found that Eureka and 1998~VF$_{31}$ are likely
Sa- or A-class asteroids, whereas 1999~UJ$_7$ is probably
an X-class asteroid.
These differing compositions suggest that
these asteroids cannot have all formed in the 
same protostellar disk environment.
\citet{rivkin07},
a companion paper,
extend the previous paper, finding that
Eureka is angritic (igneous, from an oxidized,
carbonaceous chondritic precursor), 
whereas 1998~VF$_{31}$ is likely a primitive
achondrite (from a reduced origin).
The Rivkin spectra represent the
total published knowledge of the physical properties of Mars
Trojans. Clearly, additional data are needed to
characterize these unique objects and unravel
the population's history, since Mars Trojans
may represent the only known planetesimals
that formed
interior to the asteroid belt, and 
are the closest approximations
to Earth's building blocks currently known.

We have obtained the first thermal infrared
measurements of three Mars Trojan asteroids:
(5261) Eureka, 
(101429) 1998~VF$_{31}$, and 
2001~FR$_{127}$ (this last being unconfirmed as
a Mars Trojan due to lack of long-term
dynamical integrations). Here we 
present our data (including some
ancillary visible wavelength data); thermal
modeling; and the resulting derived
diameters and 
albedos.
We briefly discuss the implications
of our results, including their general
agreement with the taxa presented
in \citet{rivkin07}.

\section{Recovery of 2001~FR$_{127}$}

2001~FR$_{127}$ was discovered in 
March, 2001, and recovered two weeks later \citep{mpec01}.
On the basis of this arc, it was identified
as a candidate Mars Trojan asteroid. (For simplicity,
we refer this body hereafter simply as a Mars Trojan.)
However,
at the time of our Gemini observations
four years later,
the positional uncertainty for this object was
around 0.67~degrees,
far too large to be
usefully targeted with Gemini/Michelle (see below),
whose field of view is less than 1~arcminute.
Before our thermal infrared observations could 
be carried out, therefore, 2001~FR$_{127}$ needed
to be recovered to reduce its positional
uncertainty.

We carried out a wide-area recovery program
in May and June, 2005, using
{\tt 90prime}, the prime focus camera
on the University of Arizona/Steward Observatory
90-inch (2.3-meter) Bok Telescope on Kitt Peak.
{\tt 90prime} has an array of four 4096$\times$4096
CCDs; the sides of the array span 1.16~deg, and 
the total imaged area of the sky is 1~deg$^2$
per pointing \citep{90prime}.
We imaged a series of overlapping fields
along the projected location of 2001~FR$_{127}$.
We used a modified version of the Deep
Ecliptic Survey data reduction and
moving object detection pipeline
\citep{millis02,elliot05}
to search
for 2001~FR$_{127}$, and found it on multiple
images. These new astrometric
positions \citep{mpec05}
allowed a substantial
refinement of the orbital elements
and a consequent reduction in positional uncertainty
for 2001~FR$_{127}$. At the time of our Gemini
observations of 2001~FR$_{127}$ one month after
recovery, the positional
uncertainty was less than an arcsecond.


\section{Partial lightcurve photometry}

%

In order to detect or place limits on any visible lightcurves,
we observed Eureka and 1998~VF$_{31}$ in V~band
(predicted magnitudes from Horizons: 18.9 and 20.6,
respectively)
on 2005~Sep~19 UT
with the facility 2K~CCD (CCD21) imager on the Steward Observatory
Mt.\ Bigelow 61-inch (1.54-meter) Kuiper Telescope.
The night was not photometric, so we cannot
independently determine the V~magnitude of the
asteroids.
Each asteroid was easily detected in a series of 
60~second exposures. 2001~FR$_{127}$ was also attempted,
but was too faint (predicted
magnitude V=22.3 from Horizons)
to be detected in 90~second exposures.
Our time baseline was quite short -- 
around 20~minutes for
each of the two detected asteroids.
No significant variation in flux from either asteroid
was detected (using relative photometry to three comparably
bright comparison stars that are nearby in the images).
For the bright Eureka, our non-detection of flux variation
places a 3$\sigma$~limit on any lightcurve variation of
0.1~mag over the 20~minute observation window.
This upper limit is consistent with the \citet{rivkin03}
measurement (from a different epoch)
of a lightcurve amplitude at least 0.15~mag
over around 6~hours. 
For the 
relatively faint 1998~VF$_{31}$, the 3$\sigma$~limit
on any lightcurve variation is around 1~magnitude over
the 20~minute observation window. Both of these non-detections
are useful in eliminating the possibility of fast
rotations and extreme shapes, which would be suggested
by large flux variations over these short time windows.

\section{Thermal infrared observations}

Mars Trojan asteroids Eureka, 1998~VF$_{31}$, and 2001~FR$_{127}$
were observed in queue mode on 2005~Jul~6 (UT) at the Gemini North
telescope on Mauna Kea using Michelle, 
a mid-infrared imager and spectrometer \citep{glasse}.
In imaging mode, Michelle has a field of view
of 32$^{\prime\prime}\times$24$^{\prime\prime}$, with
pixels 0.1~arcsec on a side.
We used the 
$N^\prime$ 
($\lambda_c = 11.2$~microns, $\Delta \lambda = 2.4$~microns)
and $Q_a$
($\lambda_c = 18.1$~microns, $\Delta \lambda = 1.9$~microns)
filters (see Table~\ref{obslog} for the observing log).
The telescope was tracked at asteroid (non-sidereal)
rates.

We employed the standard chop-nod strategy
in which the telescope secondary chops (several Hertz)
and the telescope nods (few times a minute) 
between nearby (8~arcsec) pointings in order to subtract 
out the thermal background from the telescope
and atmosphere.
The data was processed using the
{\tt midir} package of the Gemini IRAF
package\footnote{Available at
{\tt http://www.gemini.edu/sciops/data/dataSoftware.html}}.
The {\tt mireduce} task conveniently performs
all the standard tasks of reorganizing
data structures and stacking the chop-nod images
to produce final 
double-differenced
images.
In each image stack for this program only a single
target appears,
at very close to its predicted location,
so there is no confusion as to whether
the measured source is the targeted source. 
The exceptions are the $N^\prime$~image of
2001~FR$_{127}$ and the $Q_a$~image of
1998~VF$_{31}$, in which no sources are evident
at all; we derive upper limits to the fluxes for
these two observations, as described below.
In all cases, the images in the 
``off'' positions 
are unguided and somewhat
smeared, so we measure photometry
from only the ``on'' (guided)
positions.
Each ``on'' source has one half of the total
integration time.

Three photometric calibrator stars
(see \citet{cohen99}
and the Gemini web pages\footnote{{\tt http://www.gemini.edu/sciops/instruments/mir/MIRStdFluxes.html}})
were observed
using the same observing mode (Tables~\ref{obslog} and~\ref{results}).
For each standard star, we measured the total
flux using aperture photometry.
The extinction we
derived from these calibration measurements
is consistent with zero.

The Eureka and 1998~VF$_{31}$ $N^\prime$ observations
each were made with 24~total nods.
For the Eureka observations,
the image quality clearly degrades after the first
12~nods, so only these good quality
images were used.
For these data, we use a small centroided photometric
aperture of
0.5~arcsec
that is well-matched to the size
of the image. This small aperture reduces the 
sky background noise included in the photometric
aperture, but requires an aperture correction.
We derive an aperture correction of~1.62
by measuring fluxes from the standard stars
with both large and small apertures.
The final (aperture corrected) $N^\prime$ flux for
Eureka is given in Table~\ref{results}.

1998~VF$_{31}$ is somewhat fainter than Eureka,
and is not visible in data from a single nod
position. We used all 24~nod positions for this 
target, double-difference combined into three
intermediate images of eight consecutive
nod positions each.
This allows us to make three independent measurements
of the asteroid's brightness.
The final (aperture corrected) $N^\prime$ flux
for 1998~VF$_{31}$ is given in Table~\ref{results}.

At $Q_a$,
Eureka is not visible in the data from individual
nods nor easily visible in partial sums of the 
data. Consequently, we combine all the data
into a single final double-differenced image.
We used a small aperture of 0.7~arcsec, with
an aperture correction of~1.23 (derived from
the standard stars). The final (aperture corrected)
$Q_a$ flux for Eureka is given in Table~\ref{results}.

We did not detect 1998~VF$_{31}$ at
$Q_a$ band and did not detect
2001~FR$_{127}$ at $N^\prime$ band.
(Because 2001~FR$_{127}$ was not detected
at $N^\prime$, no measurement of that
asteroid at $Q_a$ was attempted.)
To determine upper limits on the fluxes
for these non-detections, we implanted
scaled copies of the $N^\prime$~1998~VF$_{31}$
point source in
the 2001~FR$_{127}$ $N^\prime$ image and
scaled copies of the $Q_a$~Eureka point source
in the 1998~VF$_{31}$ $Q_a$ image.
We measure the fluxes for the faintest
implanted objects that are detected to set
the upper limits for these
two observations.
These limits are roughly consistent with
three times the sky noise (Table~\ref{results}), as expected.

Eureka is the brightest asteroid source in our program,
and we look to it to characterize
the errors in our measurements. We have 12~independent
measurements of Eureka at $N^\prime$. 
The scatter in these flux measurements is comparable
to the variation in the sky background (both around
12\%). We therefore conclude that,
for measurements where we have only few or one
flux measurement and hence cannot characterize
the scatter in the measurements, we can estimate
the error in our measurement from the sky variability.
We use this technique for both the 1998~VF$_{31}$ $N^\prime$
and Eureka $Q_a$ measurements.

We convert Michelle counts to flux density
units (mJy) by using the photometric calibrator
stars (Table~\ref{results}).
We combine the measurements from all the 
calibrator stars to define a single calibration
factor for each bandpass
because the matches in 
time and sky location between
calibration stars and asteroids
are not very good.
At~$N^\prime$, the calibration factor
is uncertain at the 5\% level, as
estimated from the scatter among the calibration
factors from the three standard star measurements;
at~$Q_a$, the 
uncertainty, derived the same way,
is around 10\%.
We add these errors in quadrature with the errors
in our measurements to derive the final errors
reported in Table~\ref{results}.

Our targets have temperatures around
250~K (as calculated in our thermal
modeling, described below). Thus, color
corrections can be important, as these
asteroids are substantially colder than
the calibrator stars ($\sim$4000~K).
The isophotal wavelengths (wavelength at
which the flux density from the object's spectrum
equals the average flux density calculated
by integrating over the filter profile)
for Ceres (217~K)
in Michelle $N^\prime$ and $Q_a$ are
11.52~microns and 18.26~microns, respectively
(K. Volk, priv.\ comm.).
We adopt these Ceres isophotal wavelengths
for our observed asteroids as our derived temperatures
are quite close to the Ceres temperature used.
We use these isophotal wavelengths
in our thermal
models described below, although
these shifts from nominal wavelengths
are relatively small and,
compared to the errors in our measurements,
unimportant.

\section{Thermal modeling and model results}


We interpret the observed thermal emission from our targets using the
Standard Thermal Model (STM; {\it e.g.} \citet{lebofsky89}). The STM
assumes
a non-rotating (or zero thermal-inertia) spherical asteroid:
dayside temperatures are in equilibrium with sunlight, while the
nightside temperature is zero. 
\citet{lebofsky86} found that the thermal emission from
asteroids frequently has a higher color temperature than would nominally
be predicted under the STM assumptions given above, and introduced an
empirical parameter, $\eta$, that allowed them to simultaneously model
the elevated color temperature of the thermal emission and the (known)
size of their targets (Ceres and Pallas).
The canonical value for this beaming parameter $\eta$ is~0.756,
but
recent studies ({\it e.g.}, 
\citet{harris98},
\citet{delboharris02},
\citet{fernandez03},
\citet{stans07}) find
that $\eta$ can be
significantly larger.

We rely on JPL's Horizons ephemeris
service for distance, phase angle, and absolute visual magnitude
($H_V$) information (Table~\ref{obslog}, Table~\ref{results}).
We relate diameter, albedo, and $H_V$ through

\begin{equation}
D = \frac{1329}{\sqrt{p_V}} \times 10^{-H_V/5}
\label{harris}
\end{equation}

\noindent where $p_V$ is the visible geometric albedo
and $D$ is the diameter in kilometers \citep{harris98}.
We use a thermal phase coefficient
0.01~mag/deg.
We assume standard
scattering behavior for the surface in the visible, resulting in a phase
integral $q=0.39$.

\begin{figure}[p!]
\begin{center}
\vspace{-10ex}
\includegraphics[width=5in,angle=270]{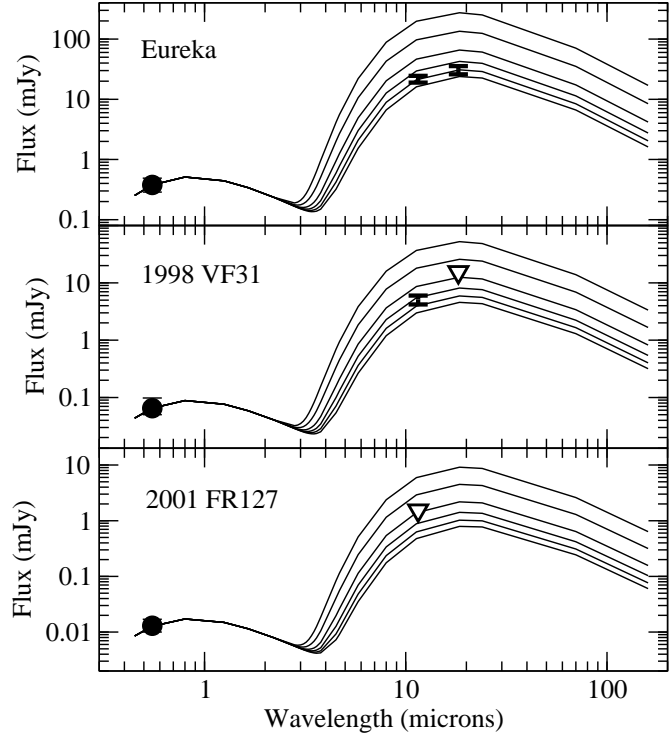}
\vspace{-5ex}
\caption[]{
Observed fluxes (plotted at isophotal
wavelengths) and
spectral energy distributions (solid curves) for the three observed
Mars Trojans (as labeled), with STM solutions for a range
of albedos and 
$\eta=1.3$.
Our detections are 
indicated by plotted error bars;
upper limits are shown as downward pointing
triangles.
The isophotal wavelengths 
for these asteroids (temperatures
around 250~K) are
11.52~and 18.26~microns 
for $N^\prime$ and
$Q_a$, respectively
(K. Volk, priv.\ comm.).
Data in the visible (circles) are
taken from JPL's Horizons service.
(Error bars on the visible fluxes,
corresponding to uncertainties
in $H$ of 0.3~magnitudes, are shown,
but are generally smaller than the symbol size.)
Based on our lack of 
detection of photometric variation (lightcurve),
we assign no extra scatter to the reflected
light (visible) data.
For each case, we 
derive diameters from thermal fluxes (assuming
$\eta=1.3$) and derive albedo from the derived
diameter and known $H_V$.
Here we show
models that correspond to six albedos:
0.05, 0.10, 0.20, 0.30, 0.40, and 0.50, from
top to bottom, respectively. 
These models assume the nominal $H$ values
for each asteroid; larger $H$ values 
imply smaller
albedos and smaller $H$ values give
larger albedos.
}
\label{sedfigure}
\label{lastfig}
\end{center}
\end{figure}

Our data on Eureka allow us to
determine the albedo $p_V$, diameter $D$, and beaming parameter $\eta$
because Eureka is strongly detected at both $N^\prime$ and $Q_a$
(in combination with $H_V$, this gives three measurements and three
unknowns). Models with $\eta=1.3$
give
the best fit
to the observed $N^\prime -  Q_a$ color (Figure~\ref{sedfigure}),
but $\eta$ values in the range~0.57--2.45
are all formally consistent with the data and error
bars given in Table~\ref{results}.
Using this range of $\eta$,
we derive a diameter for Eureka of
$1.28^{+0.44}_{-0.34}$~km. The 
corresponding albedo, when 
the diameter uncertainties and
an uncertainty of 0.3~mag for $H_V$ 
\citep{juric,rt05} are included,
is $0.39^{+0.57}_{-0.23}$, 
a range of acceptable albedos that
is so large as to be nearly useless.
We take two approaches to generating more
useful error bars.

Our first approach relies on results
from other Solar System observing programs.
Modeling of various Solar System
observations has shown that
the beaming parameter $\eta$ is unlikely to 
have a value
larger than~$\sim$1.8
\citep{fernandez03,delbo03,delbo07,stans07}. 
Requiring $\eta\leq1.8$
reduces the upper bound on 
Eureka's size and consequently the lower bound
on the albedo,
giving a diameter of $1.28^{+0.22}_{-0.34}$~km
and
an albedo of
$0.39^{+0.57}_{-0.18}$
(including the uncertainty in $H_V$).
Furthermore,
Eureka's albedo is unlikely to be larger
than~0.5 \citep{harris02,delbo03,rivkin07}.
which constrains the upper bound of the albedo
and the lower bound of the diameter, giving
$1.28^{+0.22}_{-0.29}$~km and
and an albedo of $0.39^{+0.11}_{-0.18}$.

Our second approach simply assigns
$\eta=1.3$ (the best fit for $\eta$)
rather
than using a range of $\eta$ values.
This gives a diameter of $1.28^{+0.08}_{-0.06}$~km and
a resulting albedo of 0.39$^{+0.18}_{-0.13}$
(with the uncertainty in $H_V$ still included).
We take these latter results as our 
derived best fits (Table~\ref{results}), but emphasize that
diameter and albedo both depend on choice
of $\eta$, whose uncertainty is not captured
in the error bars for this best fit,
and that alternate assumptions produce
different ``best'' results (as in our first
approach).
However, we note that in all cases the
derived (and best-fit)
albedo is consistent with
the asteroid's S(I)~taxonomic class
and interpreted angritic composition
\citep{rivkin07}.

Drawing upon the success of the STM in fitting
the Eureka data, we apply the STM
with $\eta=1.3$ to the data for 1998~VF$_{31}$
and 2001~FR$_{127}$; the results are shown
in Figure~\ref{sedfigure} and Table~\ref{results}.
This extrapolation of the STM to these 
other two bodies is warranted
both because the sizes and temperatures
of these three asteroids do not differ
significantly,
and because the
available data for 1998~VF$_{31}$ and
2001~FR$_{127}$ are so sparse that they
do not justify a different model.
Because we have only a single mid-infrared
data point for 1998~VF$_{31}$,
we solve for the albedo and diameter (Table~\ref{results})
simply
by finding the STM solution that passes through
the $N^\prime$~measurement.
The errors on diameter
give the range of solutions
that are consistent with the 1$\sigma$ photometric
error bars; the albedo uncertainties include
both the diameter uncertainty and 0.3~magnitudes
uncertainty in $H_V$.
For 2001~FR$_{127}$,
we solve for the
maximum radius
and minimum albedo
by finding the STM solution that passes through
the $N^\prime$~upper limit measurement.
The lower limit on albedo here includes an
uncertainty of 0.3~magnitudes
for $H_V$.


\subsection{Eureka's thin regolith}


\citet{spencer89} define the dimensionless
thermal parameter $\Theta$, which gives
the ratio of the characteristic radiation
timescale to the diurnal (rotation)
timescale:

\begin{equation}
\Theta = 
\Gamma \left[ \frac{\left( \omega R^3 \right)^2}{\epsilon \sigma \left[\left(1-A\right)S_1\right]^3} \right] ^ \frac{1}{4}
\label{thermal}
\end{equation}

\noindent where
$\Gamma$ is the thermal inertia of the asteroid (units of 
J/m$^2$/K/sec$^{1/2}$);
$\omega$ is the rotational frequency ($\omega = 2*\pi/P$, where
$P$ is the rotational period);
$R$ is the heliocentric distance to the asteroid, in AU (Table~\ref{obslog});
$\epsilon$ is the emissivity (we use $\epsilon = 0.9$);
$\sigma$ is the Stefan-Boltzmann constant;
$A$ is the bolometric albedo, which
is $p_V\times q=0.137$; and
$S_1$ is the solar constant at 1~AU (we use 1366~W/m$^2$).
For small asteroids with $\eta=1.3$ and
phase angle $\sim$30~degrees, 
$\Gamma$ is expected to be 
200--400~J/m$^2$/K/sec$^{1/2}$
and
$\Theta$ is likely to 
be 1--2 \citep{delbothesis,delbo07}.

\citet{rivkin03} 
show a partial lightcurve for Eureka
that suggests that Eureka's rotation
period may be $\sim$10~hours.
This period gives
$\Theta$ of 1.6--3.2 (for the above range
of $\Gamma$), in good agreement with
the expected value.
Thus, Eureka's thermal properties appear to
be consistent
with theoretical
expectations \citep{delbothesis} as
well as with the results for similarly-sized
near Earth objects (NEOs)
\citep{delbo07}.

In the STM, dayside temperature goes as $\eta^{-1/4}$, so
$\eta < 1$
results in warmer emission, while $\eta > 1$ results in cooler emission.
Relatively cooler dayside temperatures will be
produced by a high thermal inertia surface since
some of the thermal emission will occur on the nightside.
Our result that $\eta$ is greater than unity
is therefore consistent with our argument
that $\Gamma$ (thermal inertia) is relatively
large. Both conclusions are consistent
with a relative dearth of regolith, as predicted for bodies
of this size \citep{binzel04,cheng04}, and with the
small size we derive ({\it e.g.,} \citet{delbo07}).
We note that relatively large $\eta$ could also
indicate a relatively smooth surface (on macroscopic
scales), which again could imply a relatively thin
regolith.


\section{Summary and discussion}

We observed the Mars Trojan asteroids
Eureka, 1998~VF$_{31}$, and 2001~FR$_{127}$
at $N^\prime$ (11.2~microns) and
$Q_a$ (18.1~microns).
Using the Standard Thermal Model,
we derive diameters (albedos) of
1.28~km (0.39), 0.78~km (0.32),
and $<$0.52~km ($>$0.14), respectively. 
From several lines of argument, we conclude
that 
Eureka's regolith is likely
relatively thin,
as predicted for a body this size.
Eureka's thermal inertia is likely
similar to that for comparably-sized
NEOs.

\citet{rivkin07}
find that Eureka is angritic and that
1998~VF$_{31}$ appears to be an
S(VII) (e.g., primitive achondrite) asteroid.
In both cases,
the albedos we derive are 
consistent with those taxa and implied
compositions.

2001~FR$_{127}$
has a diameter ($<$520~m) comparable 
to very small NEOs that
have been observed in the mid-infrared
({\it e.g.}, \citet{delbo03}) and in
radar experiments
({\it e.g.}, \citet{ostro02}). These are among the smallest
objects studied in the Solar System.
By this virtue, further physical studies 
(spin, shape, lightcurve, etc.) are interesting on
their own.
Additionally,
comparing physical properties between
the dynamically old Mars Trojan asteroids
studied here
and the comparably sized but dynamically young
NEOs may be useful in understanding the 
evolution of these smallest asteroids.

\ack
We thank Marco Delb\'{o} and Josh Emery for thoughtful, 
thorough, and helpful reviews.
We thank Ed Olszewski and Grant Williams for their assistance
with making {\tt 90prime} observations.
Marc Buie developed 
the IDL moving object pipeline that was used to 
detect and measure the position for 2001~FR$_{127}$.
Larry Wasserman helped with the modification of
this pipeline for other observational platforms
and also provided useful input on positional errors
for planning the 2001~FR$_{127}$ recovery observations.
We thank Chad Engelbracht for helpful discussions
about color corrections and Kevin Volk (Gemini)
for providing the color corrections (isophotal
wavelengths) for Michelle.
We thank Tom Geballe, Scott Fisher, and Chad
Trujillo (Gemini) and Rachel Mason and Michael
Merrill (NOAO) for help with planning and carrying
out the Gemini observations.
This work is based in part on
observations obtained at the Gemini Observatory, which is operated by the
Association of Universities for Research in Astronomy, Inc., under a cooperative agreement
with the NSF on behalf of the Gemini partnership: the National Science Foundation (United
States), the Particle Physics and Astronomy Research Council (United Kingdom), the
National Research Council (Canada), CONICYT (Chile), the Australian Research Council
(Australia), CNPq (Brazil) and CONICET (Argentina).
The data presented here were obtained under
Gemini program GN-2005A-Q-47.
We used the JPL Solar System Dynamics group's Horizons tool
to plan our observations and analyze our results.
This manuscript was prepared using the
Elsevier/Icarus LaTeX template created 
by Ross Beyer et al.
Finally, we are grateful to the indigenous people
of Hawai`i for allowing astronomers to use their
sacred mountain.

\label{lastpage}



\begin{thebibliography}{}

\bibitem[Binzel {\it et al.}(2004)]{binzel04}
Binzel, R.P., Rivkin, A.S., Stuart, J.S., Harris, A.W., Bus, S.J., Burbine, T.H., 2004.
Observed spectral properties of near-Earth objects: results for population distribution, source regions, and space weathering processes.
Icarus 170, 259-294.

\bibitem[Cheng(2005)]{cheng04}
Cheng, A.F., 2004.
Collisional evolution of the asteroid belt.
Icarus 169, 357-372.

\bibitem[Cohen {\it et al.}(1999)]{cohen99}
Cohen, M., Walker, R.G., Carter, B., Hammersley, P.,
Kidger, M., Noguchi, K., 1999.
Spectral irradiance calibration in the infrared. X. A self-consistent radiometric all-sky network of absolutely calibrated stellar spectra.
Astron.\ J.\ 117, 1864-1889.

\bibitem[Delb\'{o}(2004)]{delbothesis}
Delb\'{o}, M., 2004.
The nature of near-earth asteroids from the study of their
thermal infrared emission.
Ph.D. dissertation, FU Berlin.
{\tt http://www.diss.fu-berlin.de/2004/289/indexe.html}

\bibitem[Delb\'{o} and Harris(2002)]{delboharris02}
Delb\'{o}, M. and Harris, A.W., 2002.
Physical properties of near-Earth asteroids from thermal infrared observations and thermal modeling.
Met.\ Planet.\ Sci. 37, 1929-1936.

\bibitem[Delb\'{o} {\it et al.}(2003)]{delbo03}
Delb\'{o}, M., Harris, A.W., Binzel, R.P., Pravec, P., Davies, J.K., 2003.
Keck observations of near-Earth asteroids in the thermal
infrared.
Icarus 166, 116-130.

\bibitem[Delb\'{o} {\it et al.}(2007)]{delbo07}
Delb\'{o}, M., dell'Oro, A., Harris, A.W., Mottola, S., Mueller, M., 2007.
Thermal inertia of near-Earth asteroids and implications
for the magnitude of the Yarkovsky effect.
Icarus, in press.

\bibitem[Elliot {\it et al.}(2005)]{elliot05}
Elliot, J.L., 10 colleagues, 2005.
The Deep Ecliptic Survey: A search for Kuiper Belt Objects and Centaurs. II. Dynamical classification, the Kuiper Belt plane, and the core population.
Astron.\ J.  129, 1117-1162.

\bibitem[Fern\'{a}ndez {\it et al.}(2003)]{fernandez03}
Fern\'{a}ndez, Y.R., Sheppard, S.S., Jewitt, D.C.,
2003.
The albedo distribution of Jovian Trojan asteroids.
Astron.\ J.\ 126, 1563-1574.

\bibitem[Glasse {\it et al.}(1997)]{glasse}
Glasse, A.C., Atad-Ettedgui, E.I.,
Harris, J.W., 1997.
Michelle midinfrared spectromneter and imager.
Proc.\ SPIE 2871, 1197-1203.

\bibitem[Harris(1998)]{harris98}
Harris, A.W., 1998.
A thermal model for near-Earth asteroids.
Icarus 131, 291-301.

\bibitem[Harris and Lagerros(2002)]{harris02}
Harris, A.W. and Lagerros, J. S. V., 2002.
Asteroids in th Thermal Infrared.
In: Bottke, W. F., Jr., Cellino, A., Paolicchi, P., Binzel, R. P. (Eds.),
Asteroids III, Univ. of Arizona Press, Tucson, pp.\ 205-218.

\bibitem[Hergenrother {\it et al.}(2005)]{mpec05}
Hergenrother, C.W., Trilling, D.E., Spahr, T.B.,
2005. MPEC 2005-L35.

\bibitem[Juri{\'c} et al.(2002)]{juric}
Juri{\'c}, M., 15 colleagues, 2002.
Comparison of Positions and Magnitudes of Asteroids Observed in the Sloan Digital Sky Survey with Those Predicted for Known Asteroids. Astron.\ J.\ 124, 1776-1787.

\bibitem[Lebofsky and Spencer(1989)]{lebofsky89}
Lebofsky, L.A. and Spencer, J.R., 1989.
Radiometry and thermal modeling of asteroids.
In: Binzel, R.P., Gehrels, T., Matthews, M.S. (Eds.),
Asteroids II, Univ.\ of Arizona Press, Tucson, pp.\ 128-147.

\bibitem[Lebofsky {\it et al.}(1986)]{lebofsky86}
Lebofsky, L.A., Sykes, M.V., Tedesco, E.F., Veeder, G.J., Matson, D.L., Brown, R.H., Gradie, J.C., Feierberg, M.A., Rudy, R.J., 1986.
A refined `standard' thermal model for asteroids based on observations of 1 Ceres and 2 Pallas.
Icarus 68, 239-251.

\bibitem[Millis {\it et al.}(2002)]{millis02}
Millis, R.L., Buie, M.W., Wasserman, L.H.,
Elliot, J.L., Kern, S.D., Wagner, R.M., 2002.
The Deep Ecliptic Survey: A search for Kuiper Belt Objects and Centaurs. I. Description of methods and initial results.
Astron.\ J.\ 123, 2083-2109.

\bibitem[Ostro {\it et al.}(2002)]{ostro02}
Ostro, S.J., Hudson, R.S., Benner, L.A.M., Giorgini, J.D., Magri, C., Margot, J.-L., Nolan, M.C., 2002.
Asteroid radar astronomy.
In: Bottke Jr., W.F., Cellino, A., Paolicchi, P., Binzel, R.P. (Eds.),
Asteroids III, Univ. of Arizona Press, 151-168.

\bibitem[Rivkin {\it et al.}(2003)]{rivkin03}
Rivkin, A.S., Binzel, R.P., Howell, E.S., 
Bus, S.J., Grier, J.A., 2003.
Spectroscopy and photometry of Mars
Trojans. Icarus 165, 349-354.

\bibitem[Rivkin {\it et al.}(2007)]{rivkin07}
Rivkin, A.S., Trilling, D.E., Thomas, C.A., 
DeMeo, F., Spahr, T.B., Binzel, R.P., 2007.
Composition of the L5 Mars Trojans: Neighbors,
not Siblings. Icarus, in press

\bibitem[Romanishin \& Tegler(2005)]{rt05}
Romanishin, W., Tegler, S. C., 2005.
Accurate absolute magnitudes for Kuiper belt objects and Centaurs.
Icarus, 179, 523-526.

\bibitem[Scholl {\it et al.}(2005)]{scholl05}
Scholl, H., Marzari, F., Tricarico, P., 2005.
Dynamics of Mars Trojans. Astron.\ Astroph.\ 175,
397-408.

\bibitem[Spencer {\it et al.}(1989)]{spencer89}
Spencer, J.R., Lebofsky, L.A., Sykes, M.V., 1989.
Systematic biases in radiometric diameter determinations.
Icarus 78, 337-354.

\bibitem[Stansberry {\it et al.}(2007)]{stans07}
Stansberry, J., Grundy, W., Brown, M., Cruikshank, D., Spencer, J., Trilling, D., Margot, J.-L., 2007.
Physical properties of Kuiper Belt Objects and Centaurs: Spitzer Space Telescope constraints.
In: Barucci, A., Boehnhardt, H., Cruikshank, D., Morbidello, A.\ (Eds.),
The Kuiper Belt, Univ. of Arizona Press, in press

\bibitem[Tabachnik and Evans(1999)]{te99}
Tabachnik, S., Evans, N.W., 1999.
Cartography for Martian Trojans.
Ap.\ J.\ Lett.\
517, L63-L66.

\bibitem[Tichy {\it et al.}(2001)]{mpec01}
Tichy, M., 16 colleagues, 2001.
MPEC 2001-F58.

\bibitem[Williams {\it et al.}(2004)]{90prime}
Williams, G.G., Olszewski, E., Lesser, M.P., Burge, J.H., 2004.
90prime: a prime focus imager for the Steward Observatory 90-in.\ telescope.
Proc.\ SPIE 5492, 787-798.



\end{thebibliography}



\clearpage	

\begin{table}
\begin{center}
\textbf{Observing log}
\begin{tabular}{lcccccccc}
\hline
\hline
Target          & Filter     & Obs.\ time & Exp.\ time&  Airmass & $R$ & $\Delta$ & phase \\ 
                &            & (UT)      & (sec)      &          &  (AU)    & (AU)& (deg) \\ \hline
Eureka          & $N^\prime$ & 9.1675000 & 60         &  1.33    &  1.49    & 0.67 & 34.4  \\
Eureka          & $Q_a$      & 9.5359725 & 150        &  1.44    &  1.49    & 0.67 & 34.4  \\
1998 VF$_{31}$  & $N^\prime$ & 10.783750 & 120        &  1.38    &  1.54    & 0.81 & 37.2  \\
1998 VF$_{31}$  & $Q_a$      & 11.152500 & 150        &  1.50    &  1.54    & 0.81 & 37.2  \\
2001 FR$_{127}$ & $N^\prime$ & 11.989445 & 120        &  1.21    &  1.69    & 0.93 & 31.4  \\ \hline
HD 141992       & $N^\prime$ & 8.7205560 & 15         &  1.07    &  $\cdots$ & $\cdots$ & $\cdots$ \\
HD 141992       & $Q_a$      & 8.8315280 & 45         &  1.07    &  $\cdots$ & $\cdots$ & $\cdots$ \\
HD 156283       & $N^\prime$ & 11.471250 & 15         &  1.31    &  $\cdots$ & $\cdots$ & $\cdots$ \\
HD 158899       & $N^\prime$ & 12.296390 & 15         &  1.46    &  $\cdots$ & $\cdots$ & $\cdots$ \\
HD 158899       & $Q_a$      & 12.407080 & 45         &  1.48    &  $\cdots$ & $\cdots$ & $\cdots$ \\ \hline
\end{tabular}
\caption[]{\label{obslog}
All observations were made on 2005 Jul 06 (UT), with the
exact midtimes indicated.
The target asteroids and calibrator
stars are given in the top and bottom groups, respectively.
The exposure times used to measure
object flux are listed here.
The total open-shutter times, including both
``on'' and ``off'' (unguided) images, were
240~and 300~sec for asteroids
and 30~and 90~sec for calibration stars
at
$N^\prime$ and $Q_a$, respectively. However,
for all
targets, only the ``on'' (guided) images were used, and
for 
Eureka, the second
half of the $N^\prime$ data was poor and not used.
}
\end{center}
\end{table}

\begin{table}
\begin{center}
\textbf{Photometry and modeling results}
\begin{tabular}{c|cccc|cc}
\hline
\hline
Target          & $H$        & $V$        & $N^\prime$ & $Q_a$        & Albedo       & Diameter       \\
                & (mag)      & (mag)      & (mJy)      & (mJy)        &              & (km)           \\ \hline
Eureka          & 16.1       & 17.51      & 21.7 (2.8) & 30.8 (4.8)   & $0.39^{+0.18}_{-0.13}$ & $1.28^{+0.08}_{-0.06}$ \\ 
1998 VF$_{31}$  & 17.4       & 19.38      & 5.1 (0.9)  & $<$15.4      & $0.32^{+0.18}_{-0.11}$ & $0.78\pm0.06$ \\ 
2001 FR$_{127}$ & 18.9       & 21.21      & $<$1.53    & $\cdots$     & $>$0.14      & $<$0.52  \\ \hline 
\end{tabular}
\caption[]{\label{results}
$H$~magnitudes;
$V$~magnitudes at time of Michelle
observations, from Horizons;
measured fluxes at the 
isophotal wavelengths of 
11.52~microns and 18.26~microns
for $N^\prime$ and $Q_a$, respectively;
and derived physical properties (using
the modified STM).
The uncertainty in $H$ (and therefore $V$)
is taken
to be 0.3~mag.
The
1$\sigma$ errors for our derived
albedos and diameters are
given.
No observation was
made of 2001~FR$_{127}$ at $Q_a$.
These STM model results use
$\eta=1.3$.
We used the following $N^\prime$
fluxes for photometric calibration:
HD~141992, 13.586~Jy;
HD~156283, 35.096~Jy;
HD~158899, 11.754~Jy.
We used the following $Q_a$ fluxes
for photometric calibration:
HD~141992, 5.362~Jy;
HD~158899, 4.650~Jy.
(Photometric calibration fluxes 
from \citet{cohen99}
and Gemini web pages.)}
\label{lasttable}
\end{center}
\end{table}

\clearpage


\end{document}